\UseRawInputEncoding
\documentclass[api,twocolumn, 10pt, showpacs, longbibliography, superscriptaddress]{revtex4-1}
\usepackage{fullpage}
\usepackage{graphicx}
\usepackage{dcolumn}
\usepackage{bm}
\usepackage{color}
\usepackage{amsmath, amsthm, amssymb}
\usepackage[modulo]{lineno}
\usepackage[usenames,dvipsnames]{xcolor}
\usepackage{ulem}
\usepackage{epstopdf}
\newcommand{\ba}{\begin{eqnarray}}
\newcommand{\ea}{\end{eqnarray}}

\usepackage{float}
\floatstyle{boxed}

\graphicspath{ {Figures/} }

\begin{document}
\title{Graphene amplifier reaches the quantum-noise limit}
\author{Kin Chung Fong}
\email{fongkc@gmail.com}
\affiliation{Raytheon BBN Technologies, Quantum Engineering and Computing Group, Cambridge, Massachusetts 02138, USA}
\date{\today}

\maketitle

To make yourself heard in a noisy environment is no easy task. An amplifier, like a megaphone, can come to your rescue by increasing your voice's volume over the background noise. Your speech can be heard clearly. This is analogous to measuring superconducting qubits. Since their energy quanta are a few orders of magnitude smaller than the thermal noise, amplifiers are necessary to boost the signal up before being registered by apparatus at room temperature. However, having a high gain from amplifiers is not nearly enough. The physical process of amplification is also subjected to fluctuations, resulting in added noise by the amplifier that can degrade the signal-to-noise ratio. For a phase-insensitive linear amplifier, the minimum amount of this added noise is half a quantum because of quantum fluctuations\cite{Caves.1982}. Employing a parametric process to achieve this fundamental limit of amplification at radio and microwave frequencies has a long history: from using variable-capacitor diodes in the 1960s to Josephson junctions in the 1980s\cite{Yurke.1987}. With high gain and low noise, modern Josephson parametric amplifiers (JPAs)\cite{Castellanos-Beltran.2007} have quickly become a must-have in laboratories\cite{Aumentado.2020}: enabling high-fidelity qubit readouts, observing quantum jumps, tracking quantum trajectory, and even searching for the rare event when the axion dark matter converts into a microwave photon under a high magnetic field.

Writing in \textit{Nature Nanotechnology}, two independent reports by Guilliam Butseraen, et. al.\cite{Butseraen.2022} and Joydip Sarkar, et. al.\cite{Sarkar.2022}, now further advance JPAs using a two-dimensional material---graphene.

\begin{figure*}[t]
\includegraphics[width=2\columnwidth]{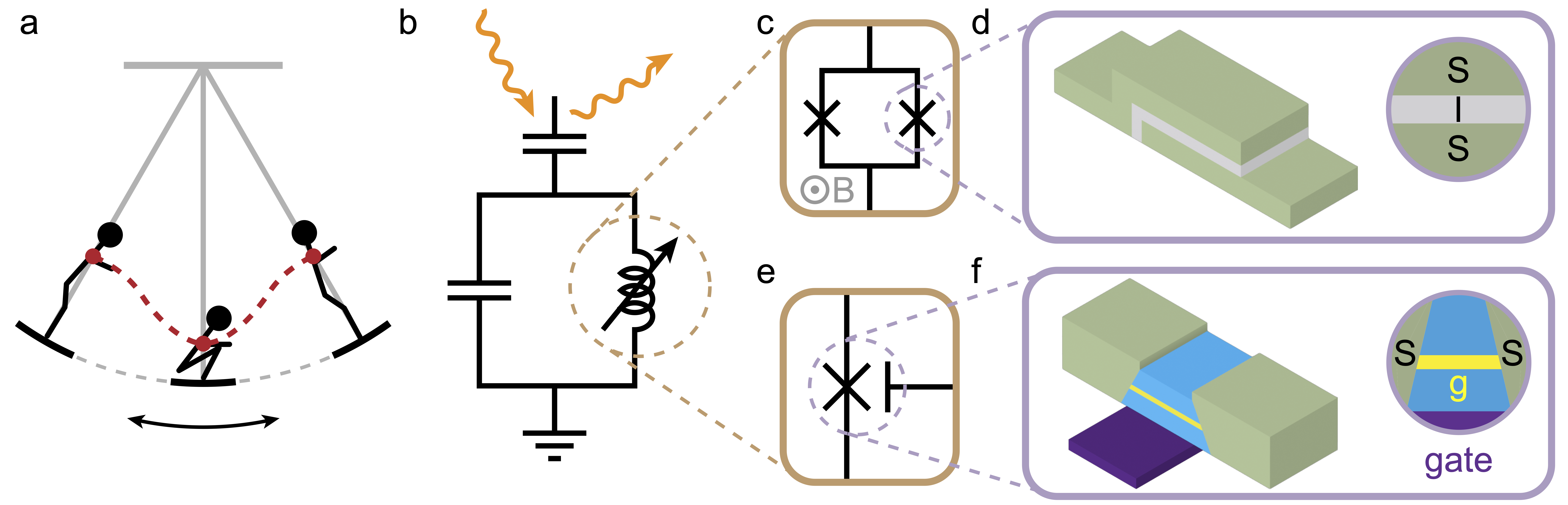}
\caption{(a) Parametric amplification in a mechanical oscillator. The red dot marks the center of mass of the child in the swing. Its oscillation periodically modulates the effective length of the pendulum, and thus, the resonant frequency. (b) Schematic model of the parametric amplifier at radio or microwave frequencies. The signal will enter the LC resonator and be amplified by the modulation of the inductance before leaving the resonator. (c) A SQUID as an implementation of the variable inductor. The Josephson inductance depends inversely on the total critical current, which is controlled by the magnetic flux through the loop. (d) Josephson junctions based on superconductor-insulator-superconductor (SIS) heterostructure. (e) Graphene-based Josephson junction with a gate control can operate as a variable inductor. (f) Josephson junctions based on superconductor-normal metal-superconductor (SNS) lateral junctions. The latest reports demonstrate JPA by SNS junction with graphene, encapsulated in hexagonal-boron nitride (blue), as the weak link. This new JPA is voltage-tunable and can operate under a high-magnetic field.}
\label{fig:GJJJPA}
\end{figure*}

We can consult children on swings about the process of parametric amplification\cite{Rugar.1991yrx} (Fig. 1a). A child can amplify the pendulum oscillation by standing up and squatting down when the swing reaches its maximum and minimum height, respectively. This stand-and-squat action is pumping  the pendulum motion at twice its resonant frequency, and doing work on the harmonic oscillator. The amplitude of the originally small oscillation (signal) will gradually increase, i.e. amplify. Parametric amplification sets apart from other amplification mechanisms because it modulates the reactance, rather than the resistance of the system. As such, it minimizes the noise that is inevitably brought into the amplification process according to the fluctuation-dissipation theorem.

A JPA constructed by a superconducting LC resonator, schematically shown in Fig. 1b, operates under the similar principle. In addition to suppressing dissipation, superconductors can form Josephson junctions, which can provide an inductance---from the inertia of the Cooper pairs tunneling through the junction---to the circuit. When two Josephson junctions are made in the form of a loop, their supercurrents will interfere and form a device known as superconducting quantum interference device (SQUID) (Fig. 1c-d). Due to the Aharonov-Bohm phase, the total critical current of the SQUID depends on the magnetic flux through the loop. A magnetic field generated by an electrical current can control the Josephson inductance. Hence, we can pump the JPA by modulating its resonant frequency with a pump current at twice the frequency of the superconducting resonators. Alternatively, JPAs can be powered up by feeding the pump tone directly into the input port for parametric amplification. The second method exploits the non-linear dependence of Josephson inductance on the magnitude of the current running through the junction, which allow for non-degenerate parametric amplifications. JPAs operate as a reflection amplifier: when the minute signal enters the JPA via the coupling capacitor, it is parametrically amplified before making its way back to the input port. 

The two research teams now replace the insulator traditionally used as the Josephson weak link with a layer of graphene encapsulated by hexagonal-boron nitride (Fig. 1e-f). By doing so, the researchers can employ the gate voltage response of the graphene to tune the Josephson critical current and thus the Josephson inductance. Hence, they can control the resonant frequency of the JPA by voltage, rather than current. Butseraen, et. al. attain parametric amplification using a transmission-line resonator at $\sim$5 GHz and a pump tone through the gate; whereas Sarkar, et. al. exploit lump components and a pump tone applied at the input port, respectively. They both achieve similar figures of merit: $\sim$10 MHz bandwidth, $\sim$500 MHz tuning of the resonant frequency with a gain of $>$20 dB, and 1-dB compression point at about -127 dBm of input power. Most importantly, both teams show that the added noise from the graphene JPA can reach the quantum limit. This is a pleasant surprise given that graphene Josephson junctions are more dissipative with a lower quality-factor at microwave frequency\cite{Haller.2022}, than the conventional superconductor-insulator-superconductor (SIS) junctions. With their careful designs and implementations, the two teams demonstrate that a superconductor-normal metal-superconductor (SNS) junction can overcome its intrinsic dissipation to achieve quantum-limited amplification.

Given the ubiquity of JPAs based on SQUIDs, it is natural to question the utility of graphene JPA beyond mere scientific curiosity. One of the potential benefits may lie in the mitigation of cross-talks among different tunable components in the quantum circuitry due to the magnetic field generated by a current bias. Without the SQUID loop, graphene JPAs can also be more densely packed and operate under a larger in-plane magnetic field. Moreover, these latest results exemplify the renewed interest in applying SNS junctions to quantum science. Unlike their insulating counterpart, SNS junctions possess more variety in their physical properties, arising from the interplay of the superconductor and the materials in the weak link. Not only would it enable voltage control of quantum amplifiers as shown here, but, by developing this motif, we can also create new hybrid superconducting electronics\cite{Baselmans.1999}, qubits\cite{Wang.2019cjf,Hays.2021}, high-sensitivity photon detectors\cite{Walsh.2021}, and topological superconductivity which may host Majorana zero modes\cite{Fornieri.2019}. With the rise of quantum technology, we shall expect more innovations from SNS junctions to come!

The author declares no conflict of interest.

\bibliographystyle{apsrev}

\end{document}